\documentclass[12pt,english,aps,prb, showpacs,preprintnumbers,amsmath,amssymb,showkeys]{revtex4-2}
\usepackage{bm}        
\usepackage{amssymb}   
\usepackage{color}
\usepackage{units}
\usepackage{amsmath}
\usepackage{natbib}
\usepackage{esint}
\usepackage{ulem}
\usepackage[english]{babel}
\usepackage[colorlinks=true, citecolor=blue]{hyperref}
\usepackage{xcolor}
\usepackage{chemformula} 
\usepackage[T1]{fontenc} 

\begin{document}

	\title{Correlated carrier dynamics in a superconducting van der Waals heterostructure}
	\author{Prakiran Baidya}
	\affiliation{Department of Physics, Indian Institute of Science, Bangalore 560012, India}
	\author{Vivas Bagwe}
	\author{Pratap Raychaudhuri}
	\affiliation{Tata Institute of Fundamental Research, Mumbai 400005, India}
	\author{Aveek Bid}
	\email{aveekbid@iisc.ac.in}
	\affiliation{Department of Physics, Indian Institute of Science, Bangalore-560012, India}

	\begin{abstract}
		Study of Berezinskii-Kosterlitz-Thouless transitions in clean, layered two-dimensional superconductors promises to provide insight into a host of novel phenomena like re-entrant vortex-dynamics, underlying unconventional metallic phases, and topological superconductivity. In this letter, we report the study of charge carrier dynamics in a novel 2-dimensional superconducting van der Waals heterostructure comprising of monolayer \ch{MoS2} and few-layer \ch{NbSe2} ($\sim 15$~nm). Using low-frequency conductance fluctuation spectroscopy, we show that the superconducting transition in the system is percolative. We present a phenomenological picture of different phases across the transition correlating with the evaluated noise. The analysis of the higher-order statistics of fluctuation reveals non-Gaussian components around the transition indicative of long-range correlation in the system.
	\end{abstract}

	\maketitle 
	
	\section{Introduction}
	With the practical realization of graphene~\cite{novoselov2004electric}, the past decade has seen an extensive exploration of layered systems. The van der Waals heterostacking of these crystalline layered materials promises to exhibit parameter-driven exotic phenomena including topologically non-trivial states~\cite{fu2008superconducting,chu2014spin,zhou2016ising,triola2016general} and strongly correlated phases~\cite{cao2018unconventional,wang2020correlated}. An example is atomically thin superconductors in the `true' 2-dimensional (2D) limit. Contrary to the 3-dimensional superconductors, for a superconductor in the 2D limit, the transition occurs through Berezinskii-Kosterlitz-Thouless (BKT) mechanism~\cite{kosterlitz1973ordering,berezinski1973jetp}. Below a characteristic critical temperature $T_{BKT}$, the vortices and antivortices are bound -- the thermal unbinding of these pairs at $T>T_{BKT}$ gives rise to the transition from the dissipationless to a finite resistive state in the system. 
	
	There are two distinct experimental strategies employed to identify a BKT transition as one approaches it from above -- (i) measurement of the superfluid density $n_S$ which is expected to go to zero discontinuously at the transition~\cite{RevModPhys.59.1001, PhysRevLett.39.1201} and (2) extrapolation of the temperature dependence of the resistivity measured at $T>T_{BKT}$ to lower $T$  using the formalism developed by Halperin and Nelson~\cite{Halperin1979}. The first technique of looking for discontinuity in the superfluid density as a signature of BKT physics does not work well for superconductors buried inside heterostructures. The second method fails for disordered superconductors due to two reasons: (i) the presence of impurities tends to broaden the transition~\cite{Sacepe2011,PhysRevB.94.085104,PhysRevB.100.174501} and (ii) the inhomogeneities change the value of the vortex-core energy from that predicted within the 2D XY model~\cite{PhysRevLett.107.217003}. 
	
	The study of carrier dynamics through resistance fluctuation spectroscopy has emerged as a powerful alternative probe to identify BKT transitions~\cite{PhysRevB.94.085104,koushik2013correlated}. Though this technique has been employed to probe the BKT physics in thin-film superconductors, the study of fluctuation statistics is not well explored in layered systems, specifically in van der Waals heterostructures. With increasing interest in such systems as platforms for realizing low dimensional superconductivity in the clean limit and topological superconductivity, there is an urgent need for a detailed study of such systems.
	
	In a previous study, we have reported the observation of two-dimensional Ising superconductivity in a van der Waals heterostructure comprising of single-layer-\ch{MoS2} (SL-\ch{MoS2}) and bulk \ch{NbSe2}~\cite{PhysRevB.104.174510}. We established that the reduced dimensionality comes from an effective thinning of \ch{NbSe2} due to the coupling with the \ch{MoS2} layer making it a perfect example of a `buried' van der Waals superconductor. Thus, the conductance fluctuation spectroscopy technique becomes very relevant to probe the BKT physics in this system. 
	
	In this letter, we report a detailed study of the carrier dynamics of this heterostructure through low-frequency conductance fluctuation spectroscopy around the BKT transition. Through systematic measurements, we establish that superconductivity has a percolative nature. We also find proof of correlated dynamics arising from long-range interaction of vortices-antivortices near the $T_{BKT}$ establishing universal BKT nature of the superconducting transition in this system.

	\section{Device Fabrication}
	To fabricate the device, we mechanically exfoliated single-layer flakes of \ch{MoS2} from a bulk crystal~\cite{PhysRevB.104.174510}. The thickness of the flake was confirmed from optical contrast and through Raman spectroscopy. The flake was then transferred onto gold contact probes pre-patterned on \ch{hBN} substrates. Subsequently, a flake of \ch{NbSe2} exfoliated inside a glove box (with oxygen and moisture levels maintained at less than one ppm) was transferred on top of the \ch{MoS2} flake. The thickness of the \ch{NbSe2} flake was estimated from its optical contrast to be $\sim15$~nm. Before extraction from inside the glove box, the heterostack of SL-\ch{MoS2}/\ch{NbSe2} was covered by a \ch{hBN} flake of thickness $\sim30$~nm to protect it from environmental degradation. Subsequently,  the stack was annealed at $200^{\circ}$~C to increase the coupling between the layers.
	\section{Results and Discussion}
\begin{figure*}[t]
		\begin{center}
			\includegraphics[width=\columnwidth]{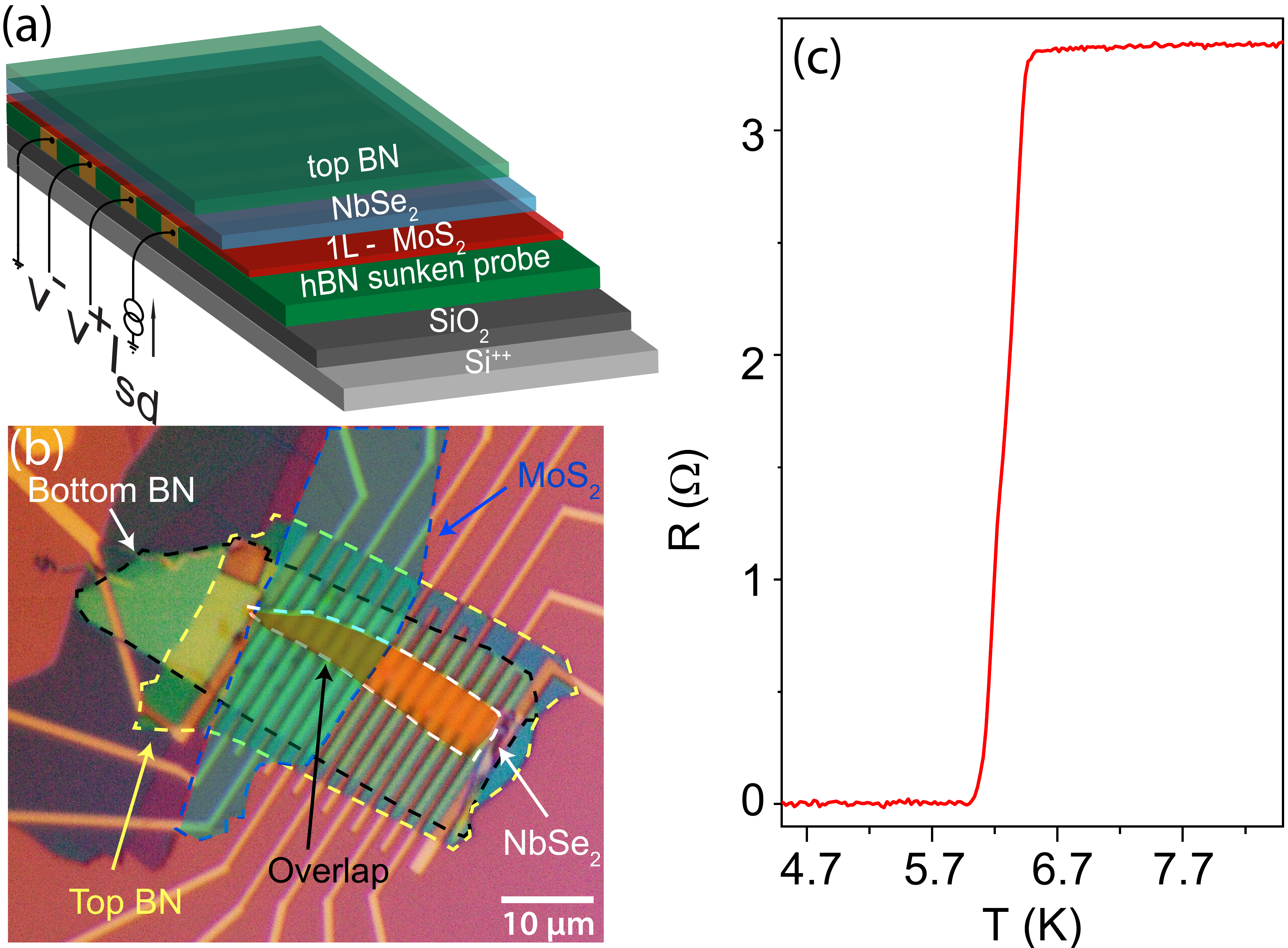}
			\caption{\small {(a) A schematic of the device structure. (b) False colour Differential Interference Contrast image of the device with different colors defining different layers of the heterostructure -- SL-\ch{MoS2}(green) and pristine \ch{NbSe2} (orange) and overlap  region (light-red). (c) Temperature dependence of the four-probe resistance of the heterostructure.}\label{Fig.1}}   
		\end{center}
	\end{figure*}

	For the initial characterization of superconducting properties of the heterostructures, electrical transport measurements were done using a DC current source and a nano voltmeter in a four-probe configuration (Fig.~\ref{Fig.1}(a)). The temperature dependence of resistance $R$ shows a metallic behavior at high temperatures followed by a transition to a superconducting state (Fig.~\ref{Fig.1}(c)) at $T^0_{C}= 6$~K. From the $R$ versus $T$ plot and also from non-linear current-voltage relations, we estimate $T_{BKT}$ to be $6.1 \pm 0.1$~K.(see Supplementary Information~\ref{Fig.S1} for details, also Ref.~[\onlinecite{PhysRevB.104.174510}]).
	
	We investigated the fluctuation statistics of the system around the $T_{BKT}$ using a 4-probe resistance fluctuation spectroscopy technique~\cite{scofield1987ac, PhysRevLett.119.226802, doi:10.1063/1.4919793} -- the details are discussed in Supplementary Information~\ref{Fig.S2}. Briefly, the device was current biased, the voltage developed across it pre-amplified and detected by a dual-phase lock-in-amplifier (LIA). The demodulated output of the LIA was digitized at a sampling rate of $1024$~points/s using a $16$ bit analog-to-digital conversion card and transferred in the computer memory for further processing. The biasing current was always maintained at a value much smaller than the critical current of the superconductor. The acquired time series of resistance fluctuations were decimated and filtered digitally to eliminate aliasing and related digital artifacts. The power spectral density of the resistance fluctuation $S_R\left(f\right)$ was then calculated over a frequency range $4$~mHz--$4$~Hz. 
	
\begin{figure*}[t]
		\begin{center}
			\includegraphics[width=\columnwidth]{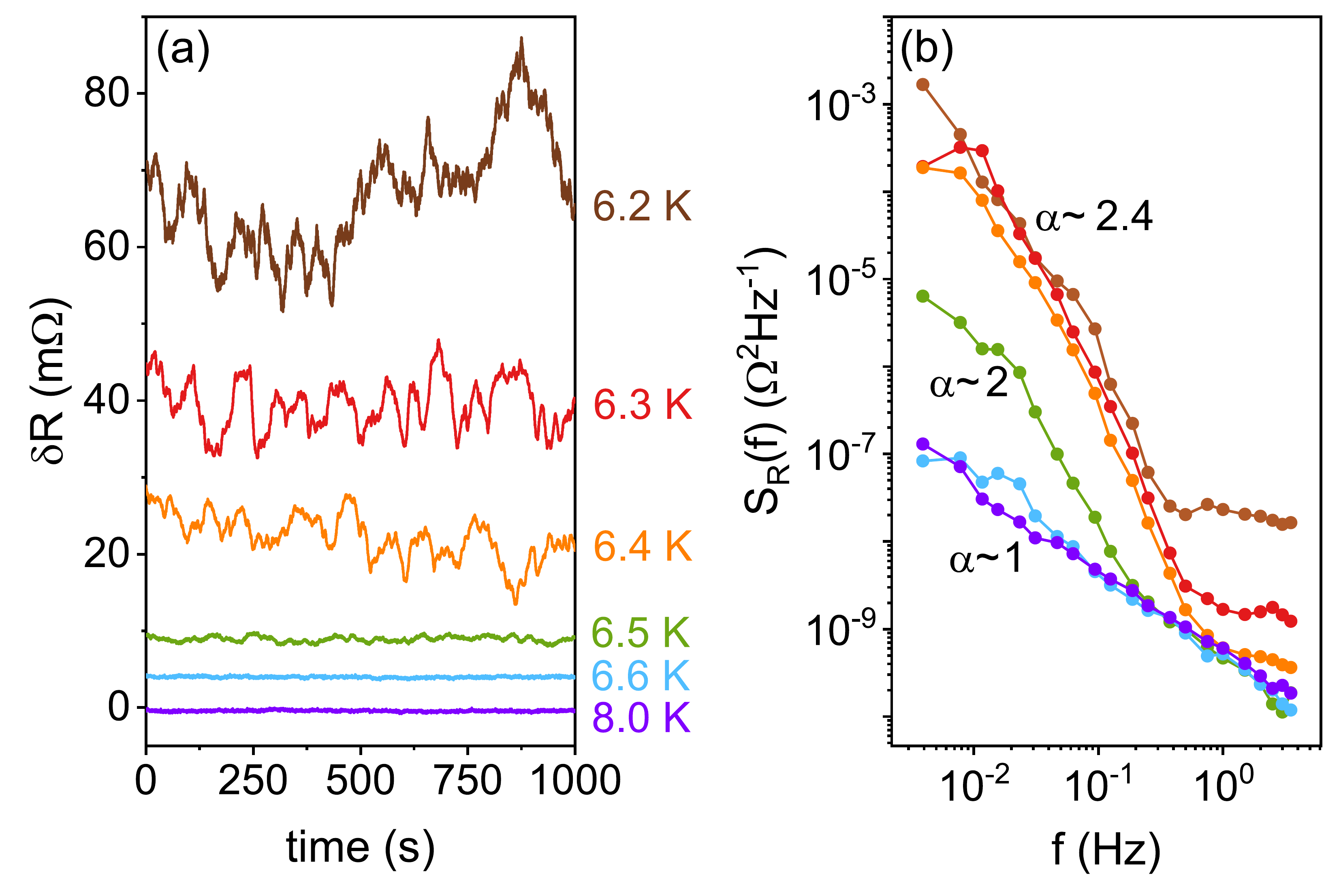}
			\caption{\small {(a) Time series of resistance fluctuations of the heterostructure region at a few representative temperatures. (b) Plots of $S_R\left(f\right)$ as function of frequency at the same values of $T$ as in (a).}\label{Fig.2}}   
		\end{center}
	\end{figure*}

\begin{figure*}
		\begin{center}
			\includegraphics[width=\columnwidth]{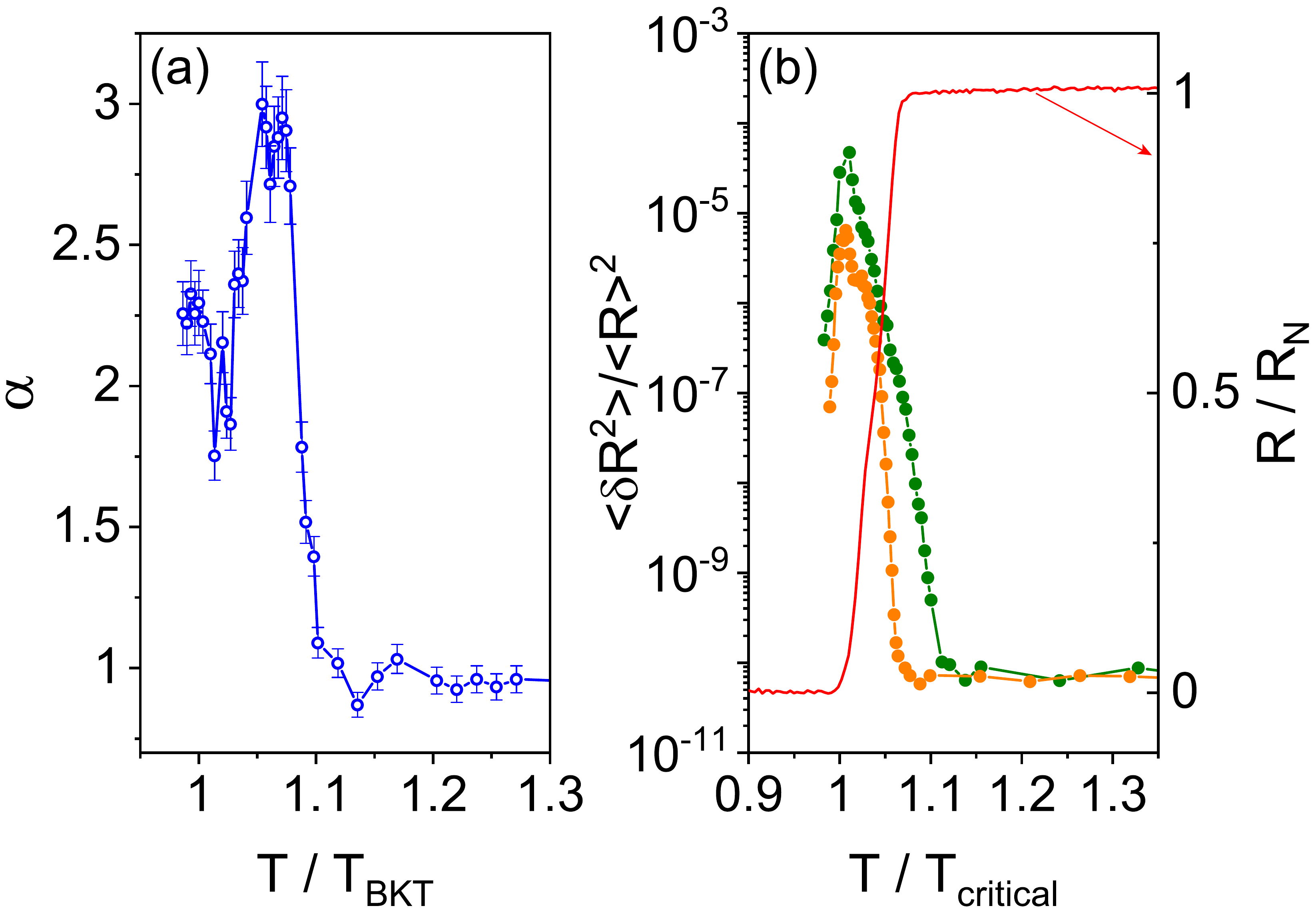}
			\caption{\small {(a) Plot of exponent $\alpha$ versus $T/T_{BKT}$ for heterostructure. (b) Plots of the relative variance of resistance fluctuations $\mathcal{R}$  for heterostructure (solid green circles) and for the pristine 3D \ch{NbSe2} region (open orange triangles) as function of $T/{T_{critical}}$. Here  $T/T_{critical}$ is $T/T_{BKT}$ for the heterostructure and $T/T^{0}_{c}$ for the  pristine \ch{NbSe2}. On the right-axis are plotted the normalized resistance $R/R_N$ for the heterostructure (red line).}\label{Fig.3}}
		\end{center}
	\end{figure*}

	Fig.~\ref{Fig.2}(a) is a plot of the time traces of resistance fluctuations for our device measured at a few representative temperatures, $T$. The fluctuations increase in amplitude with $T$ approaching $T_{BKT}$ from above. The corresponding $S_R\left(f\right)$ were found to have a frequency dependence of the form $S_R\left(f\right)\propto 1/{f^\alpha}$ (Fig.~\ref{Fig.2}(b)). One can see that the power spectral density $S_R\left(f\right)$ increases by several orders of magnitude with decreasing temperature, reflecting the observed increased resistance fluctuation in Fig.~\ref{Fig.2}(a). Additionally, the value of the exponent $\alpha$ for $f<0.5$~Hz (the method of evaluation the exponent is discussed in Supplementary Information~\ref{Fig.S3}) increases monotonically from $\sim1$ at higher temperature to $\sim2.4$ near $T_{BKT}$ (Fig.~\ref{Fig.3}(a)). There can be two possible reasons for this increase in $\alpha$ -- (i) transition of the system across different vortex phases, e.g., from an ordered to a disordered regime~\cite{jung2003noise} or (ii) fluctuation in the domain parameter of different phases across the transition range~\cite{kiss1993conductance}. Discriminating between these two scenarios requires further analysis and is beyond the scope of the current letter. 
	
	The relative variance of resistance fluctuations (we refer to this as the noise) was evaluated by integrating $S_R\left(f\right)$ over the  bandwidth of measurement~\cite{scofield1987ac, PhysRevLett.119.226802, doi:10.1063/1.4919793}:
	\begin{equation}
		\mathcal{R} = \frac{\left<\delta R^2\right>}{\left<R^2\right>}= \frac{1}{R^2}\int S_R (f)df.
		\label{eqn:1}
	\end{equation}
	Fig.~\ref{Fig.3}(b) shows the plots of relative variance and the normalized resistance against $T/T_{critical}$ ($T/T_{BKT}$) for the heterostructure region. We observe that $\mathcal{R}$ in the normal state is $\sim10^{-10}$. This value is almost five orders of magnitude lower than that reported for a typical semiconducting TMD~\cite{sarkar2019probing} attesting to the high quality of our heterostructure. 
	
	With decreasing $T$, $\mathcal{R}$ increases by nearly six orders of magnitude over a very narrow temperature window near $T_{BKT}$. As we discuss later, this divergence in noise can be well explained in terms of a percolation network model of superconducting fluctuations~\cite{PhysRevB.94.085104}. Moreover according to the percolation model in the transition regime the system should follow the relation, $S_R\left(f\right)/R^2~\propto R^{-l_{rs}}$ where $l_{rs}$ is the percolation exponent which takes up the value $\sim0.9$ in the classical picture~\cite{kogan2008electronic}. In our case, the exponent $l_{rs}$ comes out to be $0.89\pm0.03$ (see Supplementary Information~\ref{Fig.S4}), establishing the percolative nature of the system. Notably, when compared with that of the pristine \ch{NbSe2}, the noise in the heterostructure is almost an order higher around the respective transition temperatures (c.f. Fig.~\ref{Fig.3}(b)). 
	
	Before we proceed further, the effect of thermal fluctuations on the measured noise needs to be considered. $dR/dT$ diverges close to the critical temperature for a superconductor. Consequently, any minor fluctuation in temperature can give rise to large resistance fluctuations near $T_{BKT}$. To eliminate this trivial effect as the the origin of the large resistance fluctuations seen in our device, we evaluated the relative contribution of temperature fluctuations to the noise using the relation $\left[\left(dR/{dT}\right)\times\left(\delta T/R\right)\right]^2$.  Here $\delta T$ is the temperature fluctuation in the measurement system, which has been measured to be $<5$~mK in our case. The evaluated value of this relative variance at $T_{BKT}$ is $\sim 10^{-7}$ (see Supplementary Information~\ref{Fig.S5}). This value is at least two orders of magnitude smaller than the measured $\mathcal{R}$ near $T_{BKT}$, thus ruling out any significant contributions of temperature fluctuations in the measured noise.        
	
\begin{figure*}[t]
		\begin{center}
			\includegraphics[width=0.75\columnwidth]{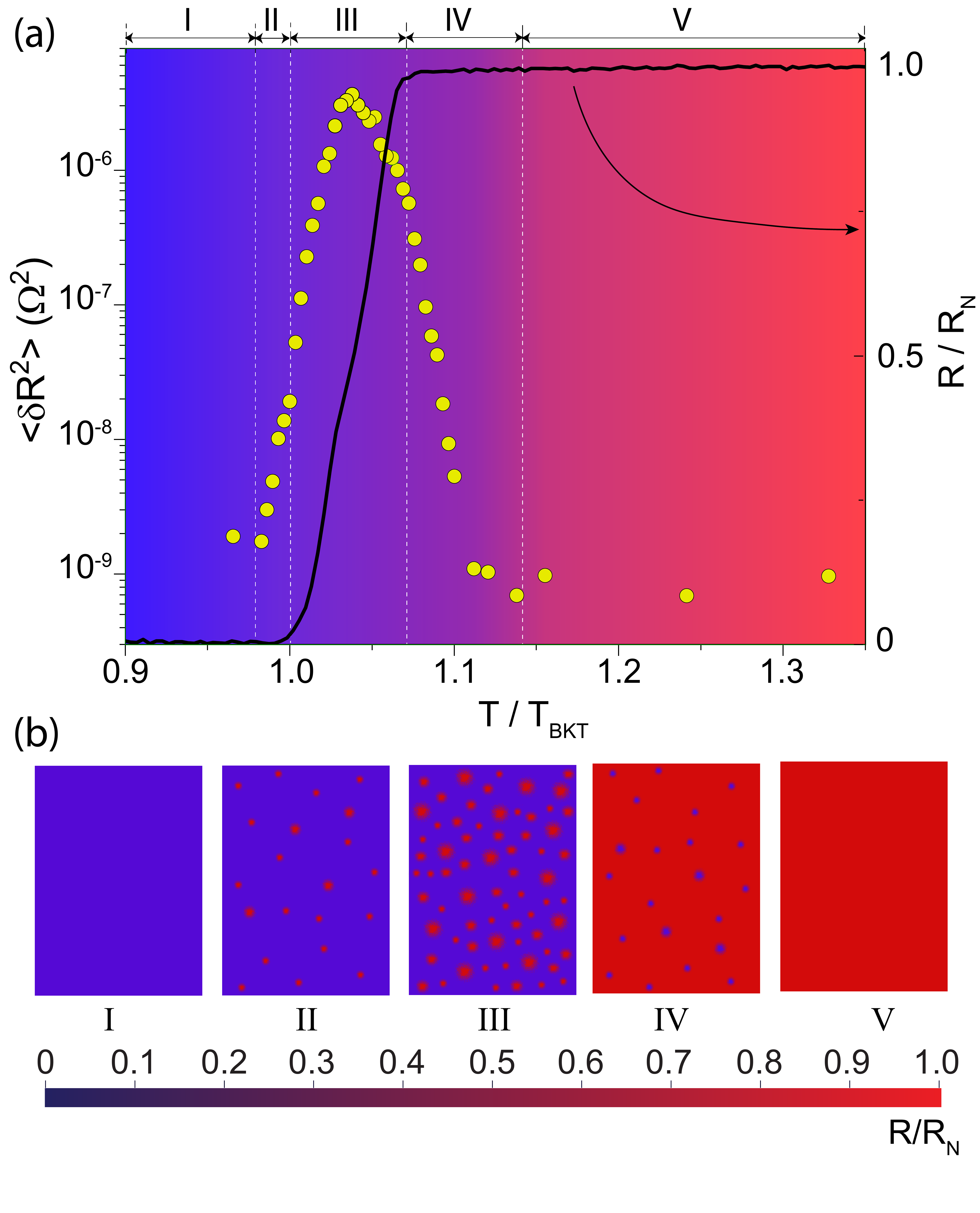}
			\caption{\small {(a) Plots of the variance of noise $\left<\delta R^2\right>$ (left-axis, yellow filled circles) and of the normalized resistance (right-axis, black solid line) as a function of $T/{T_{BKT}}$ for the heterostructure region. The color gradient indicates the transition from superconducting (blue) to normal state (red). (b) Schematics representing the microscopic status of the system in each electronic phase (for details see text). The color bar gives the value of $R/R_N$ -- blue represents the zero resistance i.e. superconducting state and red represents the normal state.}\label{Fig.4}}   
		\end{center}
	\end{figure*}

	In Fig.~\ref{Fig.4} we present a phenomenological explanation of the effect of percolation dynamics on the resistance fluctuations in a 2D superconductor in terms of a percolation network model of superconducting fluctuations~\cite{PhysRevB.94.085104}. The squares below the plot show the microscopic status of the system schematically in terms of a superconducting-normal network in different $T$-regimes. 
	
	Region-I is a purely superconducting phase. On approaching $T_{BKT}$ from below, small patches of dissipative (metallic) domains begin nucleating in the superconducting background (region-II). With increasing temperature, fluctuations in the superconducting order parameter result in the formation of a dynamic network of interconnected superconducting and normal (dissipative) regions~\cite{kogan2008electronic}. This effect is especially severe in the case of 2D superconductors. The enhanced noise in this $T$-regime has two major components -- (i) resistance fluctuations in dissipative regions; (ii) fluctuations in the number/size of the superconducting clusters~\cite{testa19881,bei2000size}. Beyond $T=T_{BKT}$, the system crosses into region III, where the proportion of superconducting and non-superconducting domains become almost equal. At this temperature (which we denote as $T_{max}$), the resistance of the device is nearly half of the normal state resistance, and the resistance fluctuation is at its maximum. The other boundary of region III comes at $T_{BCS}$, which is at $\sim6.5$~K for the system. For $T>T_{BCS}$, the fraction of the superconducting phase decreases sharply with increasing $T$ till the entire system becomes dissipative. Consequent to this decrease in electronic phase segregation of the system, the variance of resistance fluctuations decreases as the system approaches the metallic phase.             
	
	We turn now to the nature of the correlations between the fluctuating entities at different $T$-ranges. In 2D superconductors undergoing BKT transition, the XY model predicts the fluctuations to be non-Gaussian around $T_{BKT}$~\cite{bramwell2001magnetic}. These non-Gaussian resistance fluctuations have emerged as a unique signature of BKT physics and have successfully been used to discriminate between 2D and 3D superconductors~\cite{koushik2013correlated, PhysRevB.94.085104}. We quantify the non-Gaussianity of the resistance fluctuations through their  `Second spectrum,' which is the four-point correlation function of  $\delta R$, calculated over a frequency octave ($f_l,f_h$). Being extremely sensitive to the presence of non-Gaussian components (NGC), this parameter is a highly effective tool to probe correlation in a system~\cite{restle1985non, seidler1996dynamical}. To estimate the second spectrum, repeated measurements of the PSD, $S_R\left(f\right)$ is done over a selected frequency range ($f_l,f_h$). The power spectrum of this series over a frequency octave gives the second spectrum~\cite{restle1985non,seidler1996dynamical}: 
	\begin{equation}
		\begin{split}
			S^{f_1}_R\left(f_2\right) = \int_{0}^{\infty}\left<\delta R^2\left(t\right)\delta R^2\left(t+\tau\right)\right>\cos{\left(2\pi f_2\tau\right)}d\tau,\\
			\sigma^{\left(2\right)} = {\int_{0}^{{f_l}-{f_h}}S^{f_1}_R\left(f_2\right)df_2} \Big/{\left[\int_{f_l}^{f_h}S_R\left(f\right)df\right]}^2 
		\end{split}
		\label{eqn:2}
	\end{equation}    
	Here $f_1$ is the center frequency of the chosen octave and $f_2$ is the spectral frequency. $\sigma^{\left(2\right)}$ is the normalized second spectrum;  it equals 3 for Gaussian fluctuations.	
	
\begin{figure*}[t]
		\begin{center}
			\includegraphics[width=\columnwidth]{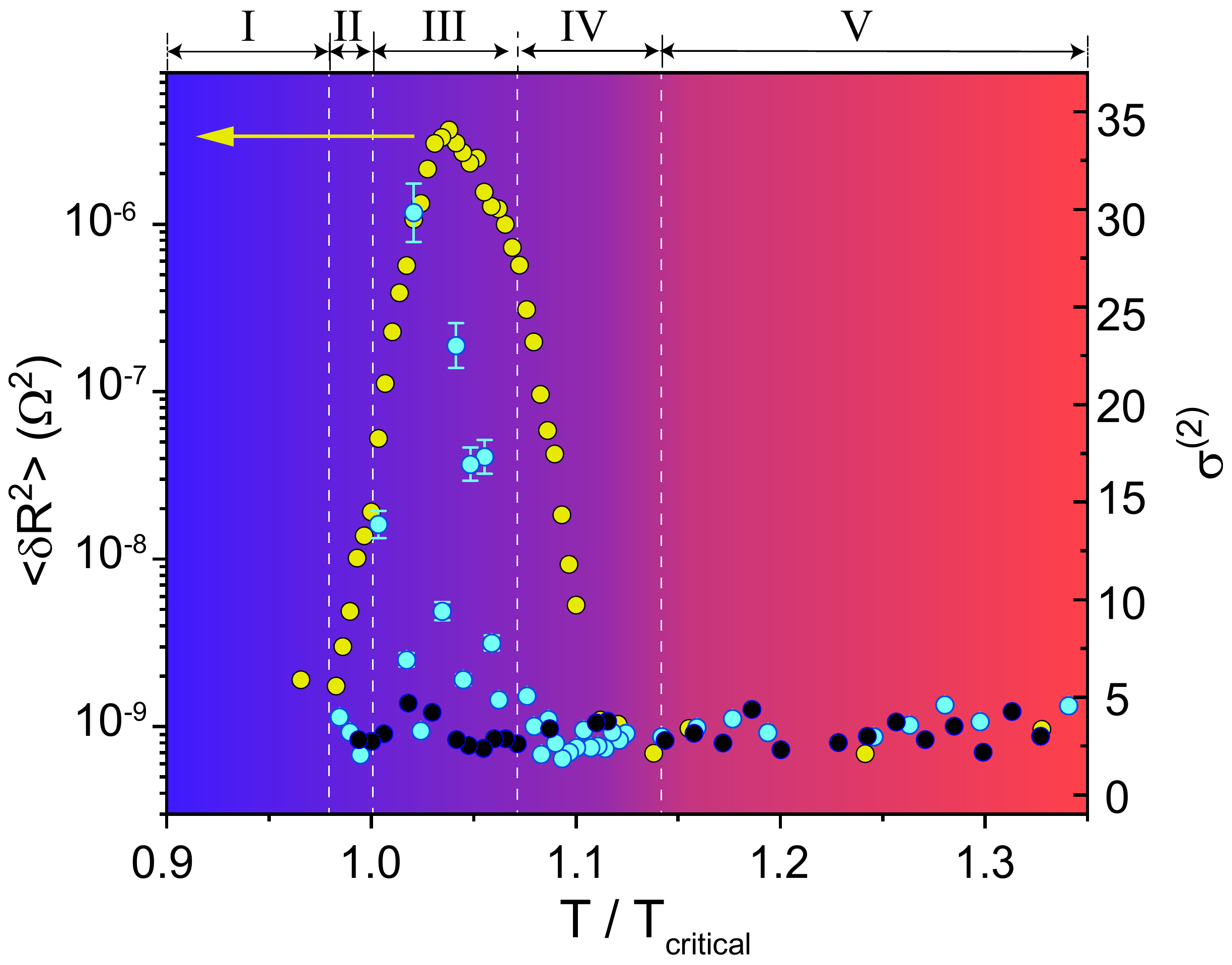}
			\caption{\small {Plots of the variance of resistance fluctuations $\left<\delta R^2\right>$ (filled yellow circles, left axis) and the normalized second spectrum $\sigma^{\left(2\right)}$ for heterostructure (filled blue circles, right axis) and for pristine 3D \ch{NbSe2} (filled black circles, right axis) as a function of $T/T_{critical}$. Here  $T/T_{critical}$ is $T/T_{BKT}$ for the heterostructure and $T/T^{0}_{c}$ for the  pristine \ch{NbSe2}. The color gradient has the same connotation as in Fig.~\ref{Fig.4}.  (for details see text).}\label{Fig.5}}   
		\end{center}
	\end{figure*}

	Fig.~\ref{Fig.5} shows the plot of measured $\sigma^{\left(2\right)}$ as a function of $T/{T_{BKT}}$ for the heterostructure region. As can be observed,  while decreasing the temperature, $\sigma^{\left(2\right)}$ increases from a baseline value of $\sim3$ at higher $T$ to $\sim30$ near $T_{BKT}$ before decaying again to the Gaussian base value for $T < T_{BKT}$.   This enhancement of $\sigma^{\left(2\right)}$ in the narrow window of $T$ in Region-III establishes clearly the appearance of non-Gaussian resistance fluctuations near $T = T_{BKT}$. In contrast, for the pristine \ch{NbSe2} region $\sigma^{\left(2\right)}$ remains at the baseline value of $\sim3$ (black solid circle in Fig.~\ref{Fig.5}) throughout the temperature range around the transition indicating a Gaussian distribution of fluctuations as expected for a 3D superconductor.
	
	Non-Gaussian fluctuations in superconductors can have different origins -- (1) long-range correlation among the vortices as has been observed in previous studies~\cite{koushik2013correlated}, (2) the dominance of percolation kinetics around superconducting transition seen in inhomogeneous superconductors~\cite{PhysRevB.94.085104}, and (3) dynamic current redistribution which appears as a consequence of substantial transport inhomogeneity and large local resistivity fluctuations that translate to the necessary condition of $\delta R/R >> 1$~\cite{seidler1996dynamical}. The third cause can be immediately ruled out by noting that in our system$\delta R/R << 1$. To discriminate between the remaining two scenarios, note that a comparison of the $T$ dependence of $\mathcal{R}$ and $\sigma^{\left(2\right)}$ (Fig.~\ref{Fig.5}) reveals that significant fluctuations in the resistance  extends beyond the onset of normal state (i.e. $R/R_N = 1$) at $T_{BCS}$ of $\sim6.5$~K. In the low-temperature limit, it extends to $T < T_{BKT}$. On the other hand, the deviation from the Gaussian value in $\sigma^{\left(2\right)}$ stays confined in region III within $T_{BKT}<T<T_{BCS}$. This deviation indicates that the increase of $\sigma^{\left(2\right)}$ that marks the existence of non-Gaussian fluctuations in the fluctuation is a consequence of an independent process which is unlikely to be the percolation kinetics that dominates the spectrum of resistance fluctuations. This strongly suggests that the first scenario of correlated vortices is at play in inducing the non-Gaussian fluctuations in the system, as has been reported earlier for clean, homogeneous superconductors~\cite{koushik2013correlated}. Further theoretical and experimental studies are essential to establish unequivocally if this is the case.
	\section{Conclusion}
	In summary, we have studied the carrier dynamics of SL-\ch{MoS2}/\ch{NbSe2} heterostructures near the superconducting transition by probing the low-frequency conductance fluctuations of the system. The first spectrum (resistance noise) shows signatures of the percolative nature of the superconducting transition. We provide a phenomenological explanation of the different phase-space regions around the transition temperature in terms of a percolative microstructure picture and correlate the resulting fluctuations with it. Furthermore, we establish the presence of strong correlations in the system around $T_{BKT}$ arising most probably from the interacting vortices and thus established that the superconducting transition in the system is of the universal BKT type.  \\

	Acknowledgments: The authors acknowledge device fabrication facilities in NNFC, CeNSE, IISc. A.B. acknowledges funding from SERB (No. HRR/2015/000017) and DST (No. DST/SJF/PSA01/2016-17)

	\clearpage
	
	\clearpage

	\clearpage

	\clearpage
	
	\clearpage

	\clearpage
	\section*{Supplementary Information}
	
	\renewcommand{\thesection}{S\arabic{section}}
	\setcounter{section}{0}
	
	\renewcommand{\thefigure}{S\arabic{figure}}
	\setcounter{figure}{0}
	
	\renewcommand{\theequation}{S\arabic{equation}}
	\setcounter{equation}{0}
	
	\section{Evaluation of BKT transition temperature}
	
	\begin{figure}[h]
		\begin{center}
			\includegraphics[width=0.8\columnwidth]{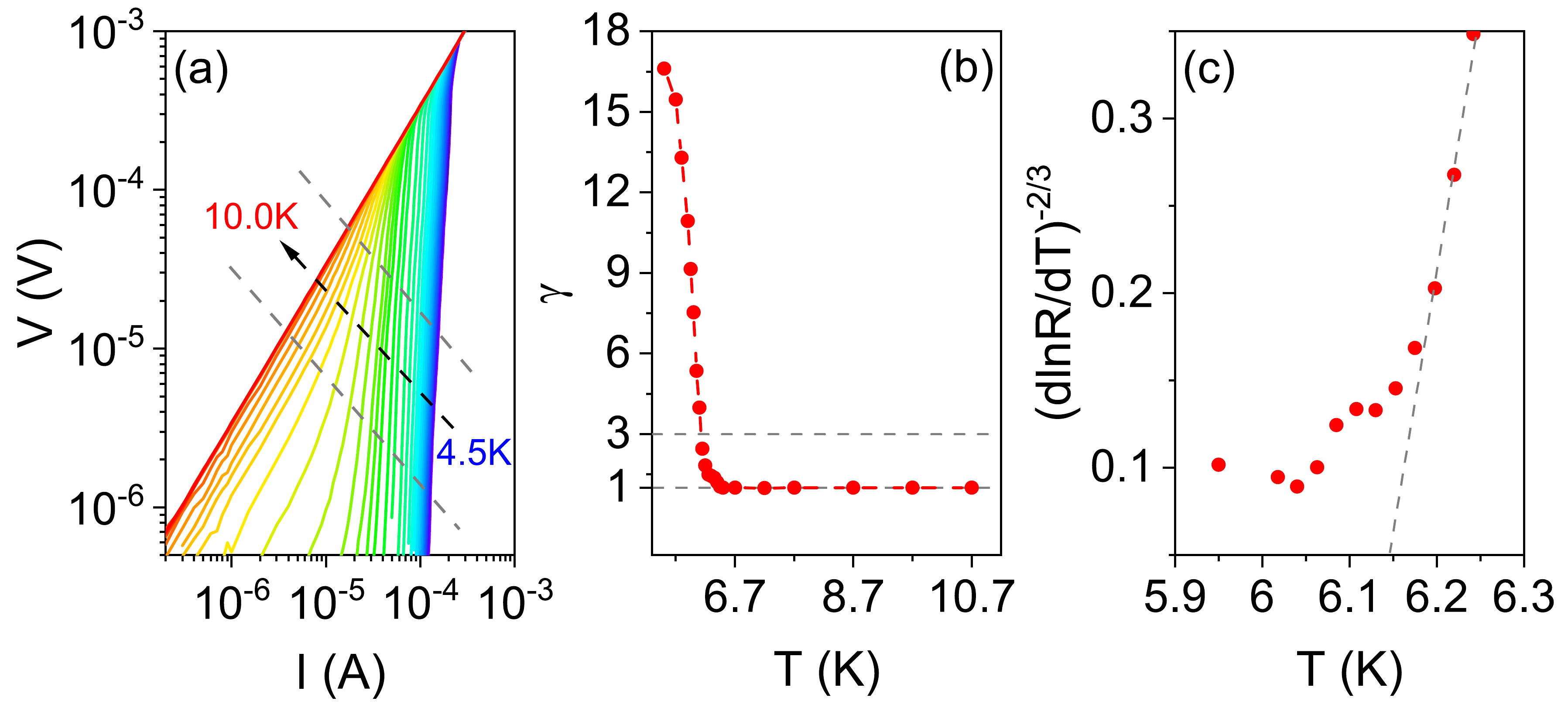}
			\caption{\small {(a) Zero magnetic field current-voltage characteristics of the heterostructure with varying temperature ranging from $4.5$~K to $10.0$~K. The red dashed line represents the range of current within which linear fit was done for each curve. (b) Plot of $\gamma$ as a function of temperature, $T$ to evaluate $T_{BKT}$ for heterostructure. (c) Plot of $(dlnR/dT)^{-{2/3}}$ as a function of temperature, $T$ for the heterostructure where the intersect of the black dashed line with the x axis gives $T_{BKT}$.}\label{Fig.S1}}   
		\end{center}
	\end{figure}
	
	For the initial characterization of superconducting properties of the heterostructure, electrical transport measurements were done using a DC current source and a nano voltmeter in a four-probe configuration.  As reported in our previous work, the superconductivity in system is of 2D nature. We thus expect the observed superconducting transition to be of Berezinskii–Kosterlitz–Thouless (BKT)  type.  For a 2D-superconductor there exists a characteristic temperature, $T_{BKT}$ below which a finite electric current can unbind the vortex-antivortex pairs system, giving rise to a dissipation which is reflected in the current-voltage characteristic as a non-linear behavior of the form, $V \propto I^{\gamma\left(T\right)}$~\cite{RevModPhys.59.1001,PhysRevB.28.2463}. $\gamma$ is a temperature dependent exponent that takes the value $3$ at $T=T_{BKT}$ and eventually goes to $1$ in the normal ohmic state. Fig.~\ref{Fig.S1}(a) shows the zero field DC non-linear current-voltage characteristics. The value of the exponent $\gamma$, evaluated through linear fitting of each curve within the marked region, are shown in Fig.~\ref{Fig.S1}(b). From this plot, we evaluate $T_{BKT}$ to be $6.13$~K. We also evaluated $T_{BCS}$ (defined as the temperature where onset of transition occurs or in other words where the IV becomes linear i.e. $\gamma = 1$) to be $6.5$~K. 
	
	The BKT transition temperature can also be obtained from resistance vs temperature plot as near $T_{BKT}$ the resistance takes the form $R = R_0\mathrm{exp}[-b_R/(T-T_{BKT})^{1/2}]$ (where $b_R$ gives the vortex-antivortex interaction strength)~\cite{RevModPhys.59.1001,ambegaokar1980dynamics,PhysRevLett.55.2156}. To evaluate $T_{BKT}$ we reduced the formula to a form, $\left(dlnR/dT\right)^{-2/3} = \left(2/b_R\right)^{2/3}\left(T-T_{BKT}\right)$. As shown in Fig.~\ref{Fig.S1}(c), the intercept of the plot of $\left(dlnR/dT\right)^{-2/3}$ vs $T$ gives $T_{BKT}$ as  to be $6.14$~K.
	
	\section{Details of noise measurement technique}
	
	\begin{figure}[h]
		\begin{center}
			\includegraphics[width=0.8\columnwidth]{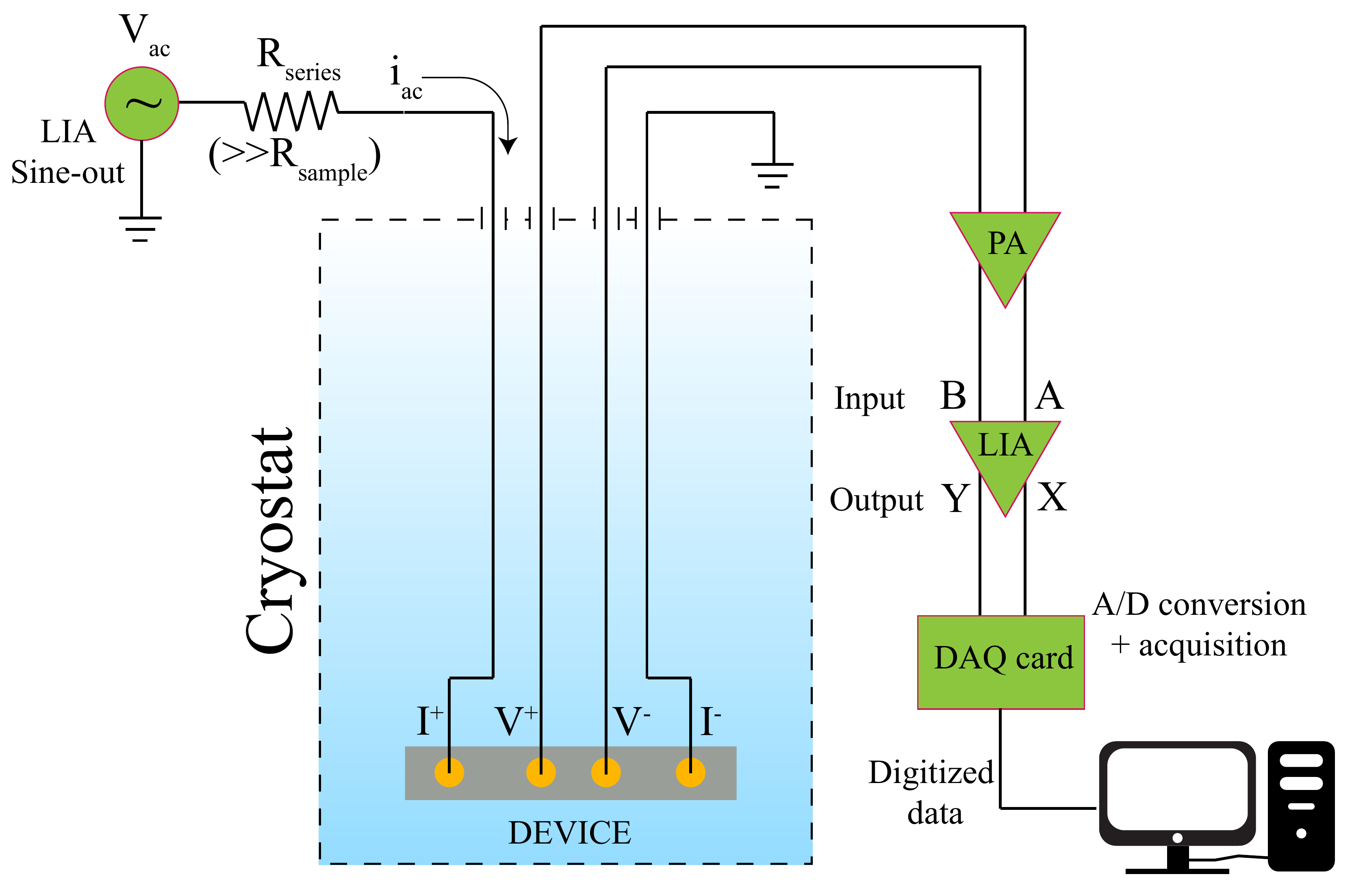}
			\caption{\small {Schematic diagram of the noise measurement setup. The sample is probed in four-probe geometry. PA represents the pre-amplifier and LIA represents the dual phase lock-in amplifier.} \label{Fig.S2}}   
		\end{center}
	\end{figure}
	
	We investigated the fluctuation statistics of the system around superconducting transition through analysis of the zero field temperature dependent resistance fluctuations acquired using an ac technique that allows us to measure the fluctuations from system and the background simultaneously~\cite{scofield1987ac}. Fig.~\ref{Fig.S2} is schematic of the measurement setup. The sample was current-biased using the sine wave output of a lock-in amplifier (SR830). A resistor, $R_{series}$ in series with the sample controls the current through it. The value of the excitation current was always maintained to be lower than the critical current of the superconductor. The voltage developed across the sample was detected using the dual-phase lock-in-amplifier coupled with a preamplifier (SR$552$). The excitation frequency of the current was kept at the eye of the noise figure of the preamplifier to minimize the contribution of amplifier noise in measured the background noise. The time constant were set to be $30$~ms with a filter roll off of $24$~dB/octave - this subsequently determines the upper cutoff frequency of the power spectral density (PSD). The output of the LIA was digitized at a sampling rate of 1024 points/s using a $16$ bit analog-to-digital conversion card and transferred in the computer memory for further processing. The in-phase channel (X-channel) picks up the excess noise from the sample as well as the background whereas the quadrature channel (Y-channel) picks up only the fluctuations from background. At every temperature, the time series of the resistance fluctuations was acquired for a duration of $30$~minutes ($\sim1.8 X 10^6$ data points). These were subsequently decimated with a factor of 128 and digitally filtered to eliminate aliasing and related digital artifacts. These filtered time series were then used to calculate the power spectral density (PSD) over the specific frequency range. The PSD of the sample noise were finally obtained by subtracting the PSD of X-channel fluctuation from that of the Y-channel.

	\section{Evaluation of the exponent $\alpha$ from $S_R\left(f\right)$}
	
	\begin{figure}[h]
		\begin{center}
			\includegraphics[width=0.7\columnwidth]{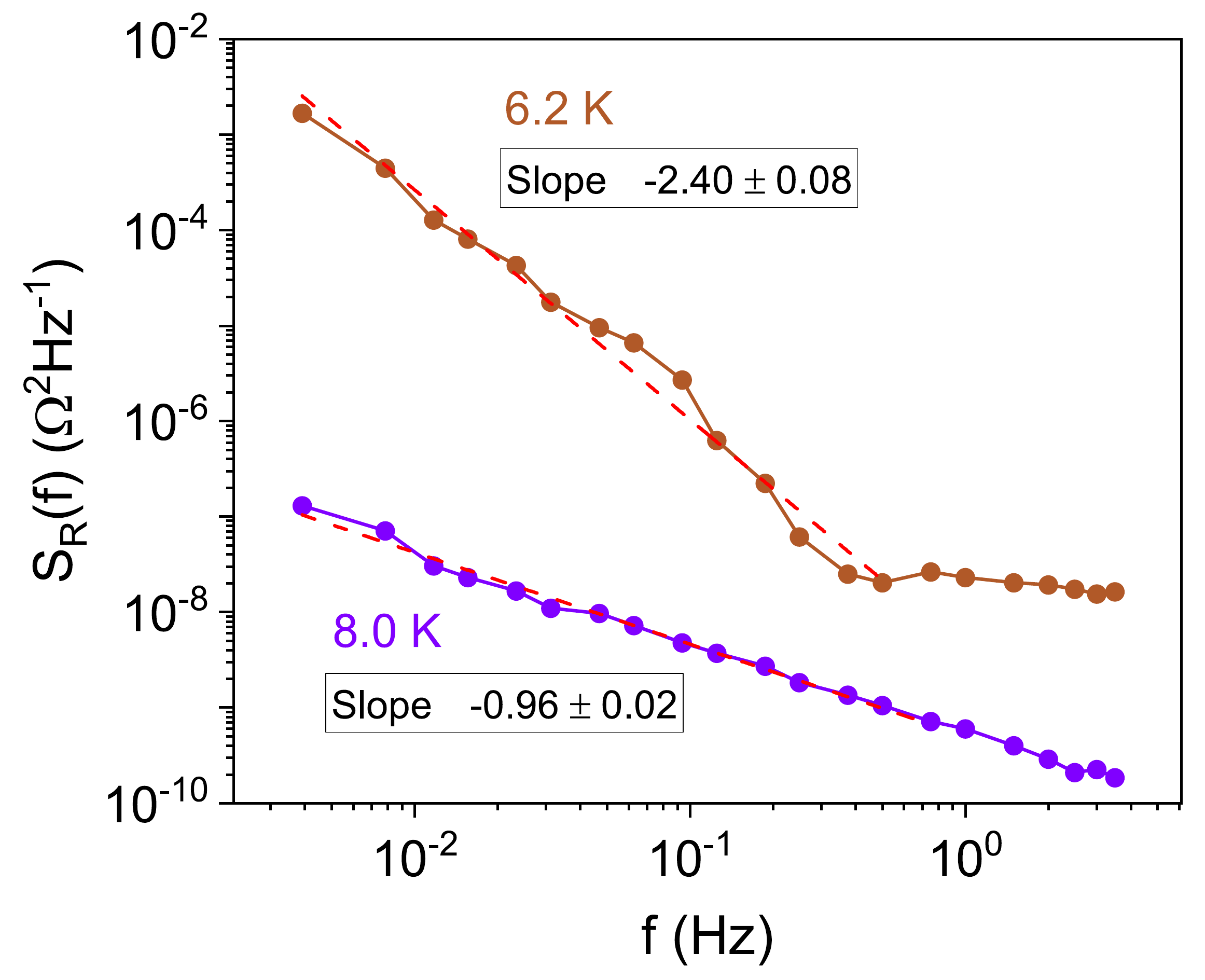}
			\caption{\small {Plot of $S_R\left(f\right)$ as a function frequency, $f$ for two different temperatures around the transition, $8$~K (violet) and $6.2$~K (brown). The dashed lines show the linear fits to the plots. The boxes with each plot show value of the slope for the respective fitting.}\label{Fig.S3}}   
		\end{center}
	\end{figure}
	
	As mentioned in the main manuscript, the power spectral density, $S_R\left(f\right)$ has a frequency dependency $S_R\left(f\right)\propto 1/{f^\alpha}$. To evaluate the exponent $\alpha$ we plotted $S_R\left(f\right)$ as function of frequency, $f$ in log-log scale, as shown in Fig.~\ref{Fig.S3}. The slope of these plots gives the value of $\alpha$. As can be seen here the slope i.e. $\alpha$ is $\sim1$ at $8$~K in which the system is in normal state whereas at $6.2$~K which is near to $T_{BKT}$ the value becomes $\sim 2.4$.
	
	\section{Classical picture of percolation}
	
	\begin{figure}[h]
		\begin{center}
			\includegraphics[width=0.8\columnwidth]{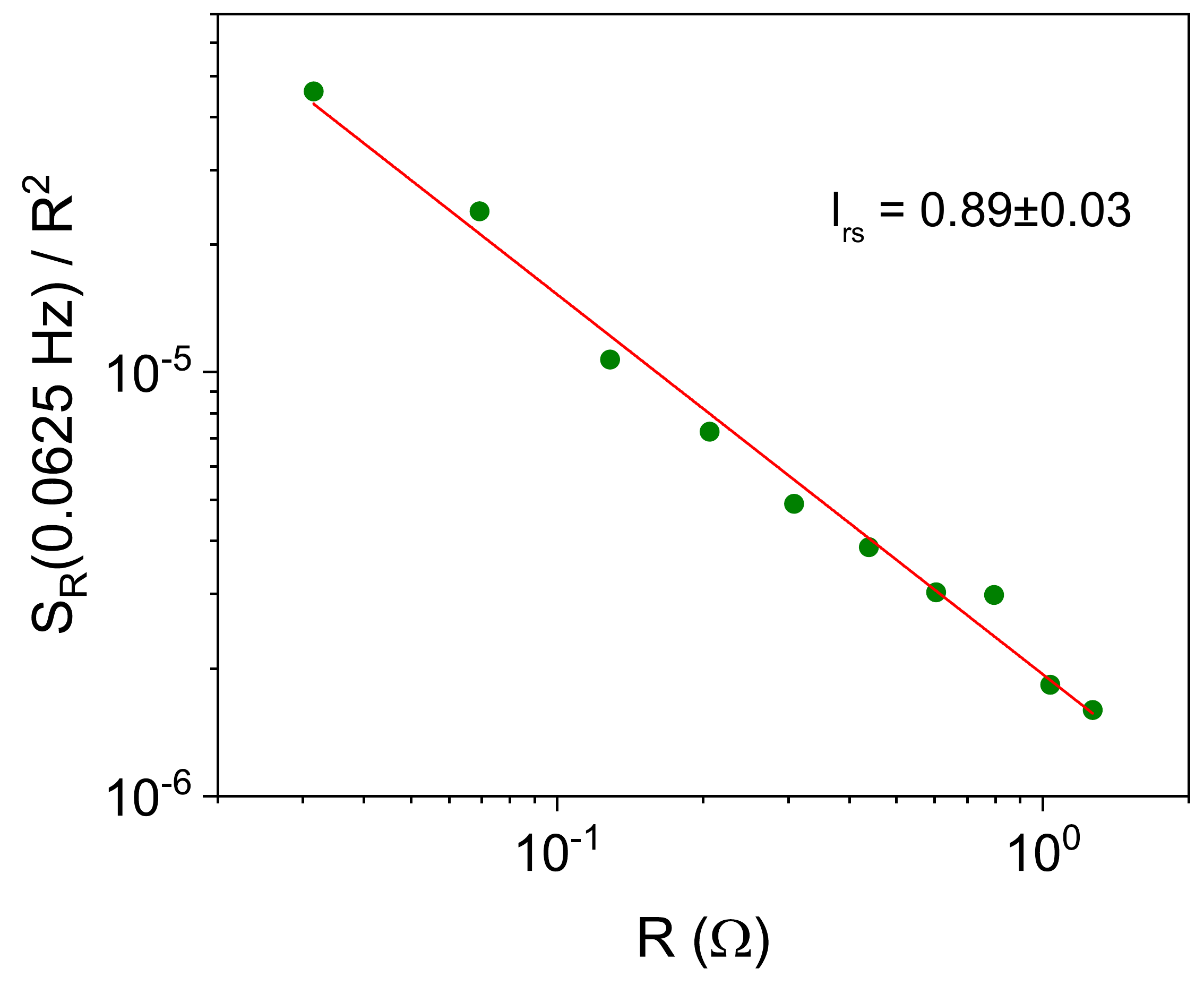}
			\caption{\small{Plot of $S_R/R^2$ at $0.0625$~Hz as function of $R$(olive solid circle). Both the axis are in log scale. The red line is the linear fit to the data that yields a slope of $\sim0.89$}
				\label{Fig.S4}}
		\end{center}
	\end{figure}
	
	For a system having percolative nature it follows that the spectral density of relative resistance fluctuation at a certain frequency, $S_R\left(f\right)/R^2$ grows as power law to the decreasing resistance towards superconductivity with a form given by~\cite{kogan2008electronic}
	
	\begin{equation}
		\frac{S_R\left(f\right)}{R^2}~\propto R^{-l_{rs}}
		\label{eqn:1}
	\end{equation}
	where $l_{rs}$ is the percolation exponent, which takes the value $\sim0.9$ in the classical picture. As can be seen in Fig.~\ref{Fig.S4} the percolation exponent for the heterostructure comes out to be $0.89\pm0.03$ which matches quite well with the classical percolation picture.
	
	\section{Contribution of temperature fluctuations to the measured noise}
	
	\begin{figure}
		\begin{center}
			\includegraphics[width=0.8\columnwidth]{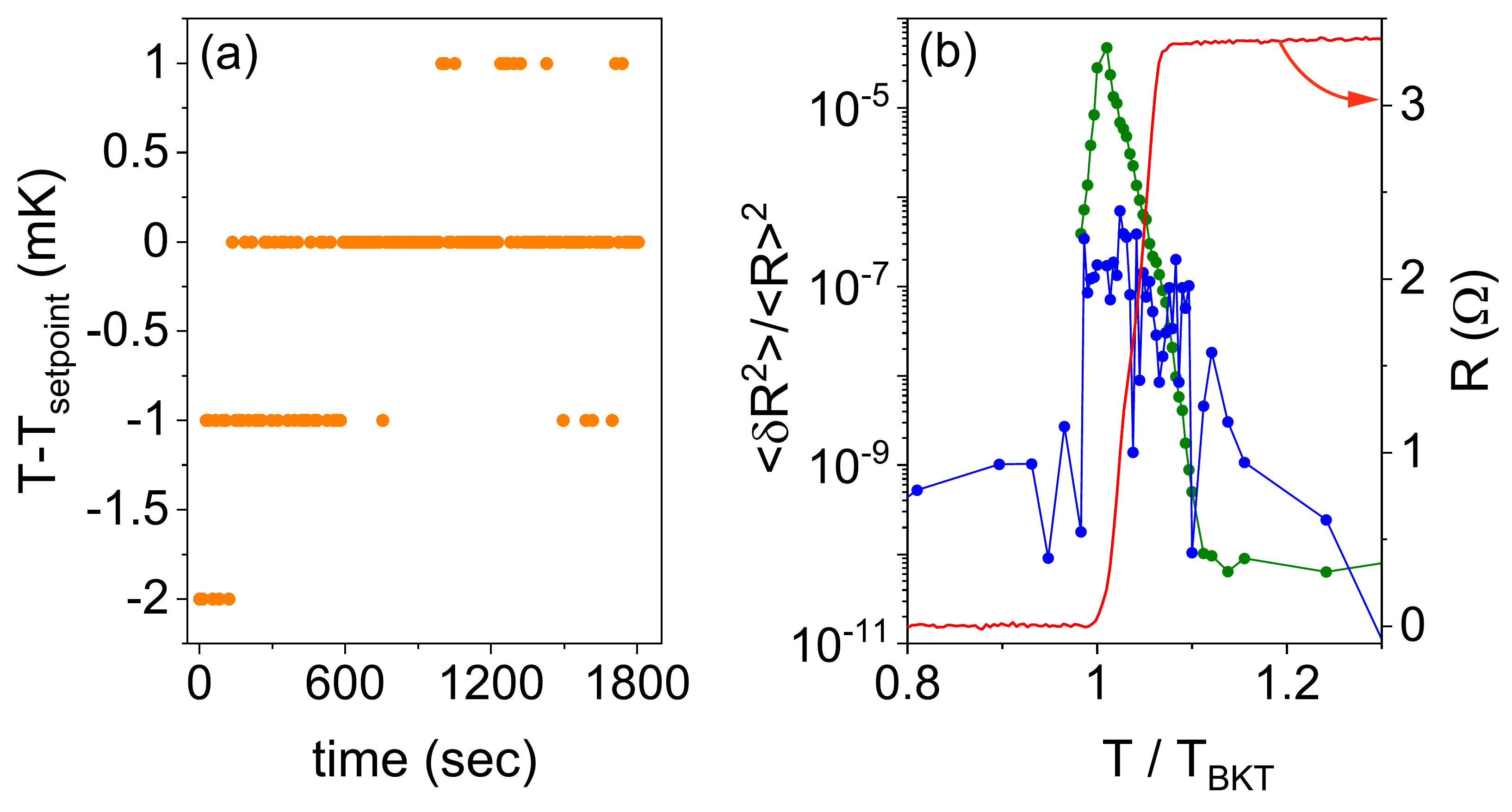}
			\caption{\small {(a) Plot of $T-T_{setpoint}$ as a function time measured at $12$~K. (b) Plot of the measured relative variance ${\left<\delta R^2\right>}/{\left<R\right>^2}$  for heterostructure (olive solid circles) and that estimated from thermal fluctuation (blue solid circles) as function of $T/{T_{BKT}}$. The lines are guide to the eye. The right-axis shows a plot of the resistance for the heterostructure region (solid red line).} \label{Fig.S5}}   
		\end{center}
	\end{figure}
	
	As mentioned in the main manuscript, we have evaluated the relative contribution of the temperature fluctuation of the measurement system to the measured noise. The temperature stability of a system depends mainly on the PID value of the temperature controller used in the experiment. We have fixed this PID value in such a way that we were able to have a temperature fluctuation, $\delta T < 5$~mK at all temperatures at which noise measurements were done. 
	
	In Fig~\ref{Fig.S5}(a), we show a plot of  $T-T_{setpoint}$ versus time over a period of 30 minutes. This is the typical time for a single noise run. Here $T_{setpoint}$ is the target temperature value (in this case, 12 K), and $T(t)$ is the instantaneous value of temperature. From Fig.~\ref{Fig.S5}(a), the maximum fluctuation is about 3 mK indicating that taking $5$~mK as $\delta T$ in our calculation is a safe choice. Fig.~\ref{Fig.S5}(b) shows the plots of the  measured relative variance (olive solid circles) and that estimated from temperature fluctuations. One can see that near $T_{BKT}$, the value of relative variance of resistance fluctuations estimated from the temperature fluctuations is almost two orders of magnitude smaller than the measured relative variance of resistance fluctuations, $\mathcal{R}$ of the heterojunction. This establishes that temperature fluctuations play a negligibly small role in the measured noise.

	\section{Noise data from another device}

	\begin{figure}[h]
		\begin{center}
			\includegraphics[width=0.8\columnwidth]{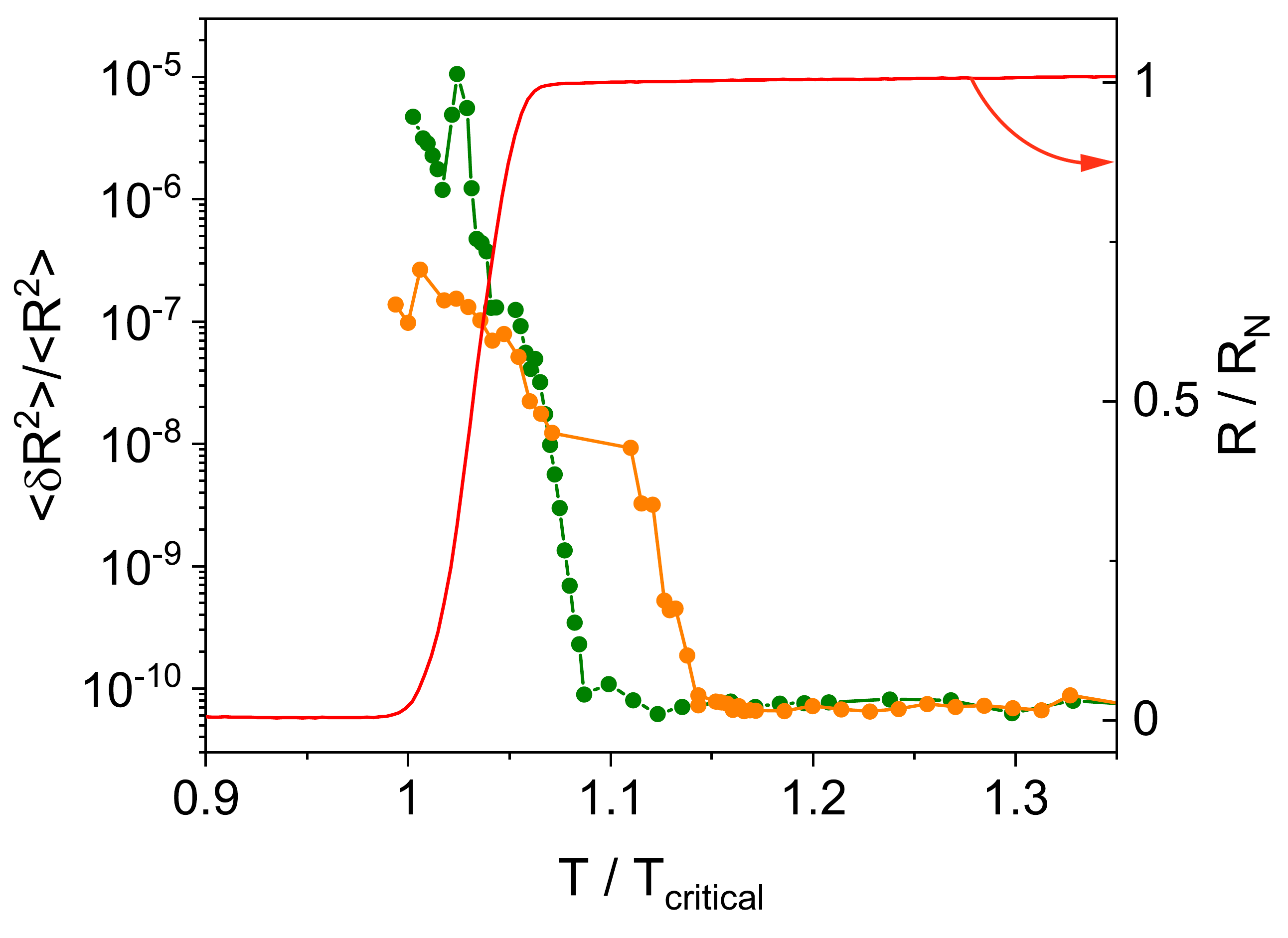}
			\caption{\small { Plots of the relative variance of resistance fluctuations, $\left<\delta R^2\right>/\left<R^2\right>$ for the heterostructure (solid olive circles) and for the pristine 3D \ch{NbSe2} region (solid orange circles) as function of $T/{T_{critical}}$ for device D2. Here  $T/T_{critical}$ is $T/T_{BKT}$ for the heterostructure and $T/T^{0}_{c}$ for the  pristine \ch{NbSe2}. On the right-axis are plotted the normalized resistance $R/R_N$ for the heterostructure (red line).}\label{Fig.S6-I}}   
		\end{center}
	\end{figure}
	
	Fig.~\ref{Fig.S6-I} shows the relative variance of resistance fluctuations of another device D2 having identical structure to the device D1 whose data were presented in the main text. As is evident from the plot, the data from D2 is very similar to that from D1 -- it shows percolative transition near $T_{BKT}$. Similar to the data for device D1, for D2 also we observe an order of magnitude higher value of the relative variance for heterostructure region in comparison to pristine \ch{NbSe2} section of the device near $T/T_{BKT}$.

	\begin{figure}[h]
		\begin{center}
			\includegraphics[width=0.8\columnwidth]{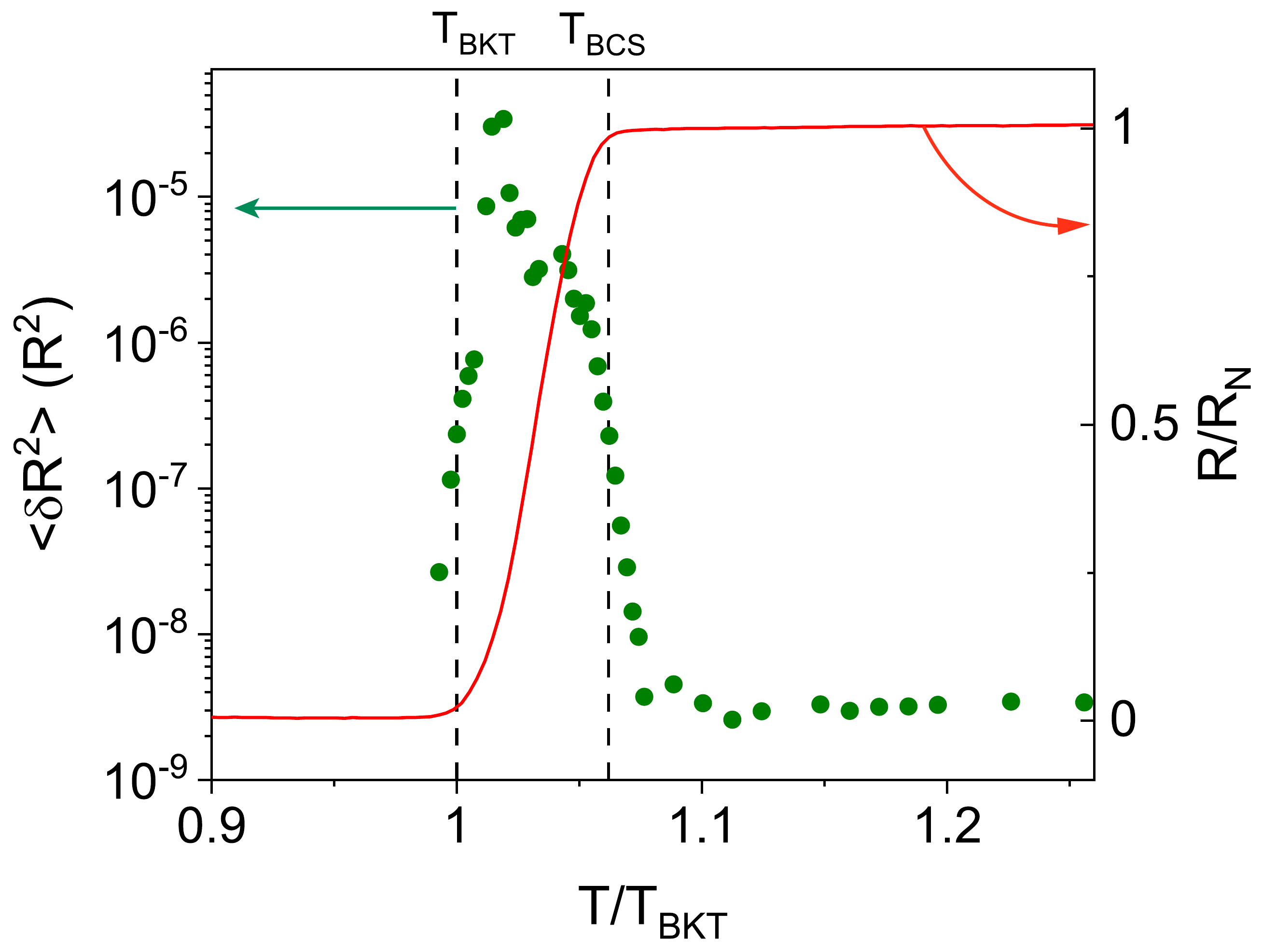}
			\caption{\small {Plots of the variance of noise $\left<\delta R^2\right>$ (left-axis, olive solid circles) and of the normalized Resistance (right-axis, red solid line) as a function of $T/{T_{BKT}}$ for the heterostructure region of device D2.}\label{Fig.S7}}   
		\end{center}
	\end{figure}
	
	Fig.~\ref{Fig.S7} shows the variance of the resistance fluctuations (olive solid circle, left axis) along with the normalized resistance, $R/R_N$ (red line, right axis)  of the heterostructure region measured for device D2 as a function of $T/T_{BKT}$. As can be seen, the increased fluctuation extends beyond $T_{BCS}$ into the normal state just as for D1 in the main text.
	
	\begin{figure}[h]
		\begin{center}
			\includegraphics[width=0.8\columnwidth]{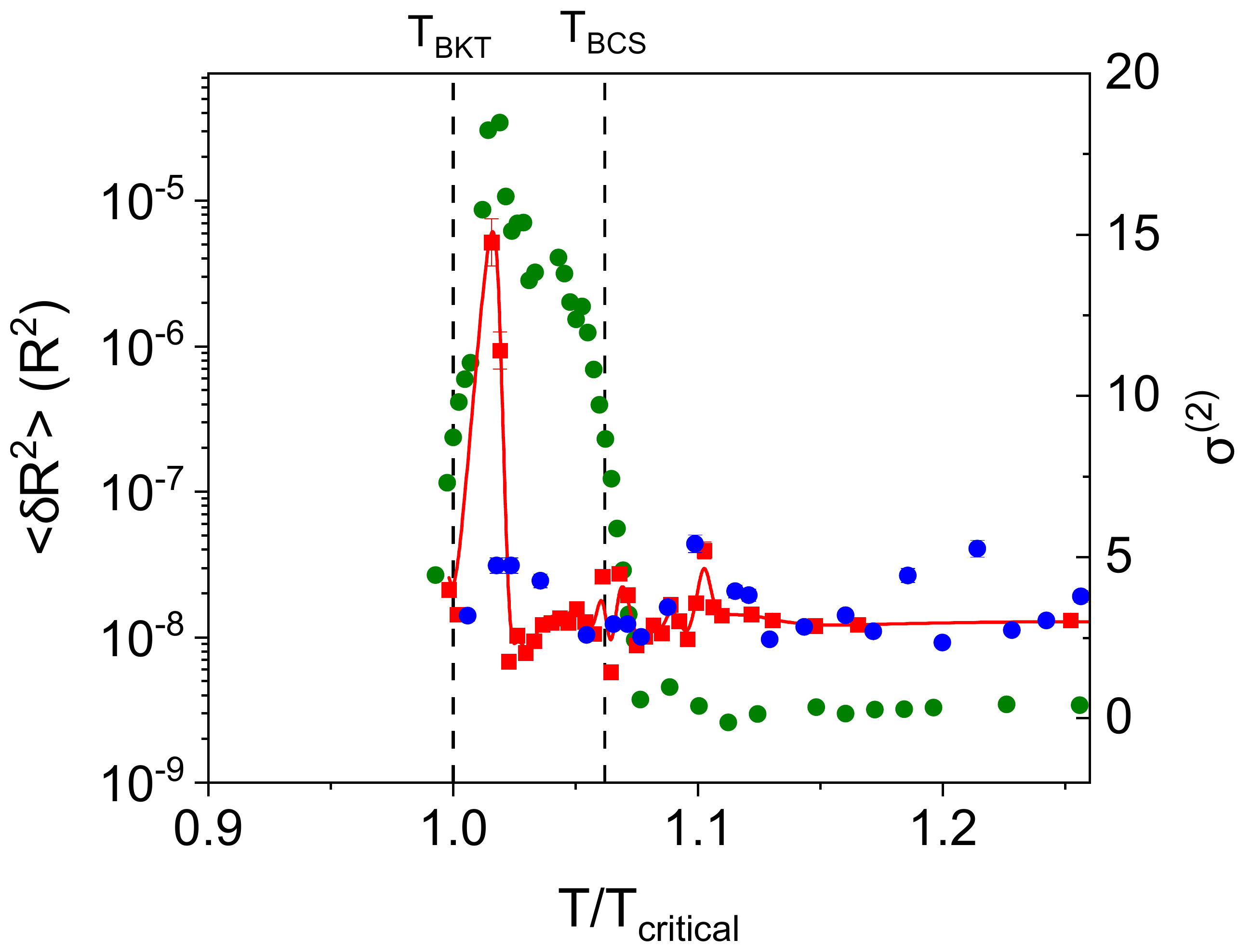}
			\caption{\small  {Plots of the variance of resistance fluctuations $\left<\delta R^2\right>$ (solid olive circles, left axis) and the normalized second spectrum $\sigma^{\left(2\right)}$ for heterostructure (solid red circles, right axis) and for pristine 3D \ch{NbSe2} (solid blue circles, right axis) as a function of $T/T_{critical}$ for device D2. Here  $T/T_{critical}$ is $T/T_{BKT}$ for the heterostructure and $T/T^{0}_{c}$ for the  pristine \ch{NbSe2}}\label{Fig.S8}}   
		\end{center}
	\end{figure}
	
	The deviation of the normalized second spectrum, $\sigma^{\left(2\right)}$ from the baseline value of $3$ in Fig.~\ref{Fig.S8} are constricted within the region bounded by $T_{BKT}$ and $T_{BCS}$ suggesting that, like device D1 in main text, the non-Gaussian nature for device D2 also occurs due to the long range correlations between the vortex-antivortex pair around the transition. Moreover, as expected we observed $\sigma^{\left(2\right)}$ to be $\sim3$ for pristine \ch{NbSe2} region of device D2 throughout the temperature range indicating the Gaussian nature of the fluctuations. This similarity in the evaluated results for the two different devices of similar heterostructure thus proves that the main observed phenomenons are inherent to the system and not device specific.

%

\end{document}